\documentclass[5p]{elsarticle}
\usepackage{amsmath,amssymb}
\usepackage{graphicx}
\usepackage{txfonts}
\usepackage[colorlinks=true]{hyperref}

\journal{Computer Physics Communications}
\begin{document}

\title{Calculation of higher-order moments by higher-order tensor
renormalization group}

\author{Satoshi Morita}
\ead{morita@issp.u-tokyo.ac.jp}
\author{Naoki Kawashima}

\address{Institute for Solid State Physics, University of Tokyo,
 Kashiwa, Chiba 277-8581, Japan}

\begin{abstract}
 A calculation method for higher-order moments of physical quantities,
including magnetization and energy, based on the higher-order tensor
renormalization group is proposed.  The physical observables are
represented by impurity tensors.  A systematic summation scheme
provides coarse-grained tensors including multiple impurities.  Our
method is compared with the Monte Carlo method on the two-dimensional
Potts model.  While the nature of the transition of the $q$-state Potts
model has been known for a long time owing to the analytical arguments,
a clear numerical confirmation has been difficult due to extremely long
correlation length in the weakly first-order transitions, e.g., for
$q=5$.  A jump of the Binder ratio precisely determines the transition
temperature.  The finite-size scaling analysis provides critical
exponents and distinguishes the weakly first-order and the continuous
transitions.
\end{abstract}


\maketitle

\section{Introduction}

Tensor networks (TN) are powerful tools for strongly correlated
many-body physics \cite{Cirac_Verstraete_review,Orus_review}.  As a
pioneering work, the density matrix renormalization group (DMRG)
\cite{DMRG1992} achieves great success in one-dimensional quantum
systems. This method can be viewed as a variational method based on the
matrix product states formalism, i.e. the one-dimensional TN states.
The projected entangled pair state (PEPS) \cite{PEPS2004} and the
projected entangled simplex state (PESS) \cite{PESS2014} provide its
generalization to higher-dimensional systems.

The tensor renormalization group method (TRG) \cite{TRG2007} provides an
efficient contraction scheme avoiding exponential divergence of the bond
dimension based on coarse-graining TNs.  Many derivatives are proposed as a
more efficient and accurate method \cite{SRG2009, SRGfull2010,
HOTRG2012, TNR2015, loopTNR2017}.  The higher-order tensor
renormalization group method (HOTRG) is a TN method applicable to a
higher-dimensional system \cite{HOTRG2012, HOTRG_3D_potts}.  The
higher-order singular value decomposition (HOSVD) provides its
truncation scheme.  An isometry which merges two bonds into one bond
plays a key role in information compression.  It is another advantage of
HOTRG than other TN renormalization methods because a local original
tensor is not decomposed into two tensors.

The partition function of a classical system can be represented by a TN
\cite{Baxter_book, SRGfull2010}.  Thus the free energy is quite
accurately estimated by using real-space renormalization group methods
mentioned above.  There are two ways to calculate an expectation value
of a physical quantity like magnetization $\langle m\rangle$.  One
method is numerical differentiation of the free energy with respect to a
model parameter, for example, the external magnetic field.  The other is
introduction of an ``impurity''; an expectation value is
represented by a TN in which one of tensors is replaced with an impurity
tensor that represents the local order parameter.

Higher moments of physical quantity are important for analysis of
critical phenomena.  Expectation value of the squared magnetization,
$\langle m^2\rangle$, is regarded as an order parameter.  The Binder
ratio is typically used to determine the transition point in the Monte
Carlo (MC) simulations \cite{Binder1981}.  Finite-size scaling analysis
of these quantities provides critical exponents.  However, calculation
methods of higher moments are limited and not accurate in TN methods.
In the numerical differentiation method, we need to calculate
higher-order derivative to estimate a higher-order moment.  Numerically
it causes cancellation of significant digits.  Especially near the phase
transition point, the higher-order derivative shows a sharp change and
it is difficult to assure its accuracy.  An efficient scheme for the TN
states was proposed based on the momentum generating
functions~\cite{West2015}.  However it also needs the numerical
differentiation with respect to an indeterminate to extract a moment.
On the other hand, the impurity method also has a problem.  Let us
assume that a system has a transitional symmetry and consider a physical
quantity represented as a spacial average of local quantities, for
example $m=(1/N)\sum_{i=1}^{N}\sigma_i$.  Only one TN which contains one
impurity tensor is sufficient to estimate its expectation value,
$\langle m\rangle$.  However, for the second-order moment, all spatial
configurations of two impurities are necessary.  Therefore, since the
number of TNs corresponding to two-point correlation functions diverges,
such an approach is unmanageable.  An exception is a one-dimensional
system, where matrix product operator representation for the high-order
moments is known~\cite{Lin2017}.

In this paper, we propose an alternative way of calculating the
higher-order moments by introducing the impurity tensors into the HOTRG
scheme.  We define a coarse-grained tensor which represents summation of
all configurations of multiple impurities.  Thus its trace directly
yields the higher-order moments.  The systematic summation technique
proposed for the infinite PEPS method \cite{Corboz2016} derives the
real-space renormalization scheme of the impurity tensor.  In contrast
to the impurity method mentioned above, the summation over impurity
configurations is taken during renormalization steps.

Using our proposed method, we investigate the ferromagnetic $q$-state
Potts model~\cite{Potts,Wu1982} on the square lattice.  A phase
transition between the ferromagnetic and paramagnetic phases is
continuous for $q\leq 4$ and discontinuous otherwise.  Especially a
numerical confirmation of the weakly first-order phase transition in the
$5$-state Potts model has attracted much attention~\cite{Ozeki2003,
iino2018}.  Since the correlation length becomes very large but finite
at the transition point, it is difficult to make a distinction between
the weakly first-order and second transitions.  However, a TN
renormalization method can easily treat the huge system size over the
correlation length.
The corner transfer matrix renormalization group method
(CTMRG)~\cite{CTMRG1996,CTMRG1997} can accurately estimate the latent
heat of the two-dimensional $5$-state Potts model~\cite{CTMRG_Potts}.
The HOTRG approach can confirm the first-order phase transition in the
$3$-state Potts model on the simple cubic lattice~\cite{HOTRG_3D_potts}.
In this paper, we present a more systematic method for characterizing
the nature of the transition.
We perform the HOTRG simulation up to the
system size $N=2^{40}$ and show that the second moment of the
magnetization clearly distinguishes between discontinuous and continuous
phase transitions.  Moreover, the finite-size scaling analysis of the
Binder ratio accurately produces the critical exponents expected in the
first-order phase transition for $q\geq 5$.

In the next section, we present the method for the higher-order moment
with a brief introduction of HOTRG.  In the third section, we perform
numerical simulations in the $q$-state Potts model on a square lattice.
We compare our method with a cluster MC method and estimate a transition
temperature from a jump of the Binder ratio.  In the fourth section, we
consider the finite-size scaling ansatz.  We estimate the critical
exponent from a slope of the Binder ratio at criticality.  The last
section is devoted to discussions and conclusions.

\section{Methods}
\label{sec:method}

In general, the partition function of a many-body system has a tensor
network representation.  For simplicity, let us consider a classical
discrete spin system with nearest neighbor interactions on the isotropic
square lattice.  A generalization to a higher dimensional system is
straightforward.  If the degrees of freedom are continuous, we use the
characterize representation.  Then we have
\begin{equation}
 Z=\sum_{\{\sigma_i\}} \prod_{\langle ij\rangle} W_{\sigma_i \sigma_j}
  \prod_{i=1}^{N} V_{\sigma_i}
  =\operatorname{tTr} \prod_{i=1}^N T_{x_iy_ix'_iy'_i},
\end{equation}
where $W_{\sigma \sigma'}$ and $V_{\sigma}$ is the local Boltzmann
factor of the nearest neighbor interaction and the magnetic field
respectively.  A variable $\sigma_i$ represents the spin degree of
freedom at a site $i$.  The local tensor $T$ locates on each lattice
site.  Based on the eigenvalue decomposition $W=\tilde{U}\Lambda
\tilde{U}^\dagger$, the local tensor always has the form
\begin{equation}
 T_{xyx'y'}^{(0)} = \sum_{\sigma}
  X_{\sigma x}X_{\sigma y}X_{\sigma x'}^*X_{\sigma y'}^*
  V_{\sigma},\label{eq:initial_T}
\end{equation}
where $X\equiv \tilde{U}\sqrt{\Lambda}$.  The superscript $(0)$
indicates a initial tensor before a renormalization.

\begin{figure}
 \centering
 \includegraphics[scale=0.45]{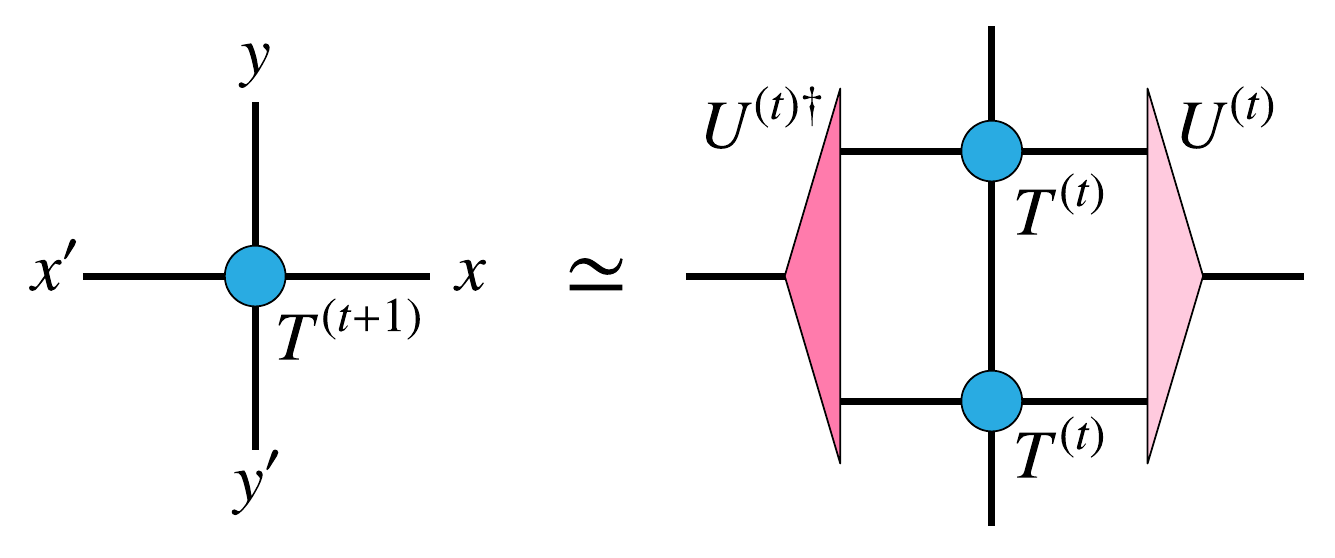}
 \caption{\label{fig:HOTRG} Graphical representations of a
 renormalization step along $y$-axis in HOTRG.}
\end{figure}

In HOTRG algorithm, the square lattice is contracted along the $x$- and
$y$-axes in sequence. The renormalization step with $y$-bond
contraction is graphically shown in Fig.\ref{fig:HOTRG}.
The renormalized local tensor at the $(t+1)$-th step is calculated as
\begin{equation}
 T_{xyx'y'}^{(t+1)} = \sum_{\substack{x_1,x_1'\\x_2,x_2',y_1}}
  T_{x_1yx_1'y_1}^{(t)}T_{x_2y_1y_2'y'}^{(t)}
  U_{x_1x_2x}^{(t)} U_{x_1'x_2'x'}^{(t)*},
\end{equation}
where $U^{(t)}$ is an isometric tensor obtained by the standard HOSVD.
In this paper, we abbreviate this contraction as
\begin{equation}
 T^{(t+1)} = \mathcal{R}(T^{(t)},T^{(t)};U^{(t)}),
\end{equation}
or, when the meaning is clear, more shortly
\begin{equation}
 T \leftarrow TT.
\end{equation}
We note that a direction of the renormalization depends on a step $t$.
The local tensor $T^{(t)}$ after the $t$-th renormalization step
represents a system of size $N=2^t$. The trace of $T^{(t)}$,
\begin{equation}
 Z \simeq \operatorname{Tr} T^{(t)} \equiv \sum_{x,y} T^{(t)}_{xyxy},
\end{equation}
yields the partition function of the corresponding finite-size system
under the periodic boundary condition.

\begin{figure}
 \centering
 \includegraphics[scale=0.35]{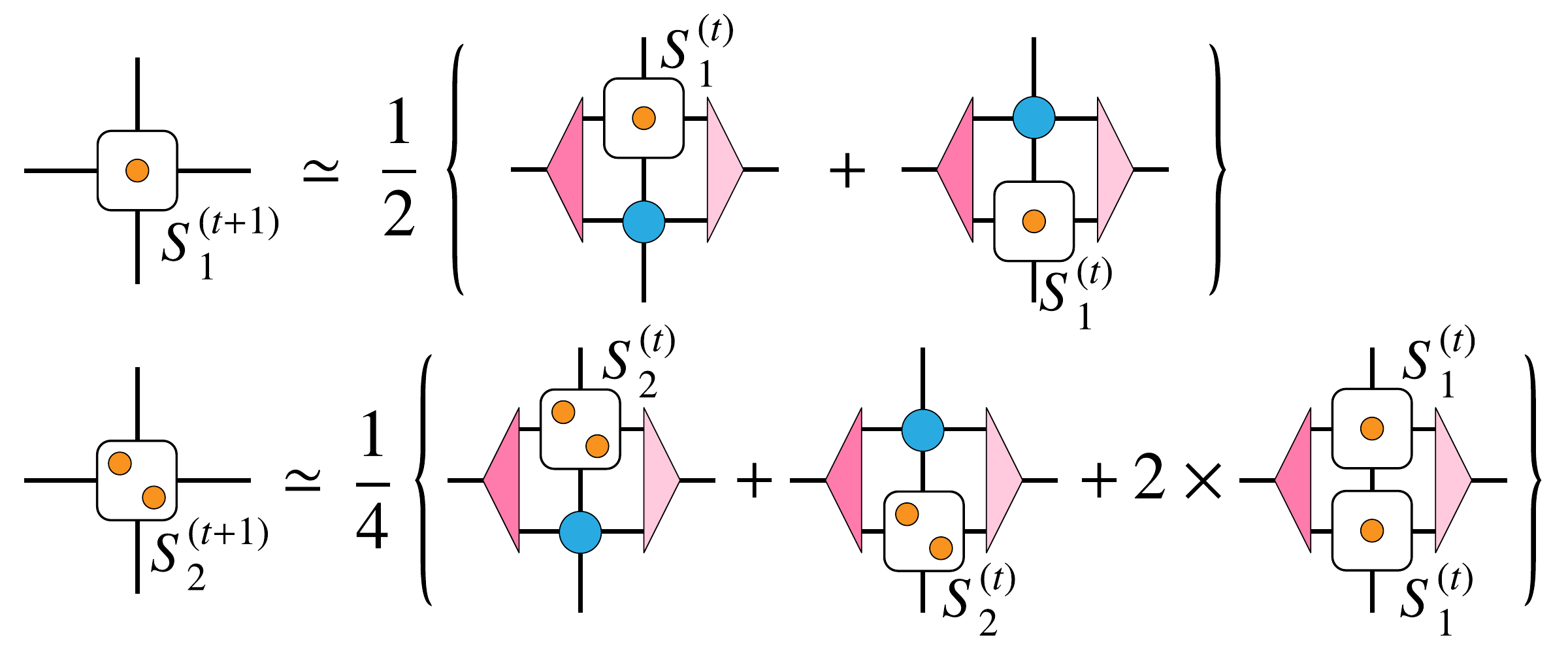}
 \caption{\label{fig:HOTRG_mag} Renormalization of the impurity tensors
 with (a) single impurity and (b) two impurities. The impurities are
 represented by a (orange) circle in the impurity tensors.}
\end{figure}

First, let us consider the magnetization defined by the average of local
magnetization as
\begin{equation}
 m \equiv \frac{1}{N}\sum_{i=1}^{N} m_{\sigma_i},\label{eq:def_m}
\end{equation}
where $m_{\sigma_i}$ represents a local magnetization at a site $i$.
With similar representation to Eq.(\ref{eq:initial_T}), the
corresponding impurity tensor is clearly obtained as
\begin{equation}
 S_{1,xyx'y'}^{(0)} =\sum_{\sigma}
  X_{\sigma x}X_{\sigma y}X_{\sigma x'}^*X_{\sigma y'}^* V_{\sigma}
  m_\sigma,
\end{equation}
where the additional subscript ``$1$'' indicates that this impurity tensor
contains one impurity in total.  In HOTRG, this single impurity tensor is
renormalized by using the same isometric tensor $U^{(t)}$ for
renormalization of $T^{(t)}$ as
\begin{equation}
 \tilde{S}_1^{(t+1)} = \mathcal{R}(S_1^{(t)},T^{(t)};U^{(t)}).
\end{equation}
The trace of this impurity tensor represents a
magnetization at a specific site but does not do the averaged
magnetization defined in Eq.(\ref{eq:def_m}). In order to finally
obtain the average over the whole lattice, we should take the local
average at each level of the renormalization procedure.  Therefore the
symmetrized version
\begin{equation}
 S_1^{(t+1)} = \frac{1}{2}
  \left\{\mathcal{R}(S_1^{(t)},T^{(t)};U^{(t)})
   +\mathcal{R}(T^{(t)},S_1^{(t)};U^{(t)})
  \right\},
\end{equation}
is required, which we write shortly as
\begin{equation}
 S_1 \leftarrow \frac{1}{2}(S_1T+TS_1).
\end{equation}
Its graphical representation is shown in Fig.~\ref{fig:HOTRG_mag}(b).
The expectation value of $m$ with the system size
$N=2^t$ under the periodic boundary condition is estimated by
\begin{equation}
 \left\langle m \right\rangle \simeq
  \frac{\operatorname{Tr} S_1^{(t)}}{\operatorname{Tr} T^{(t)}}.\label{eq:ave_m}
\end{equation}

A generalization of this scheme to higher-order moments of the
magnetization $\left\langle m^k\right\rangle$ is clear.  The initial
tensors with multiple impurities are defined as
\begin{equation}
 S_{k,xyx'y'}^{(0)} =\sum_{\sigma}
  X_{\sigma x}X_{\sigma y}X_{\sigma x'}^*X_{\sigma y'}^* V_{\sigma}
  m_\sigma^k.
\end{equation}
The update rule for the two-impurity tensor ($k=2$) can be written as
\begin{equation}
 S_2 \leftarrow \frac{1}{4} \left( S_2 T + 2 S_1 S_1 + T S_2 \right),
\end{equation}
and its diagram is shown in Fig.~\ref{fig:HOTRG_mag}(b).
Obviously $S_2$ contains all of the possible configurations of two impurities in
a finite system corresponding to the renormalization step.
The generalization to more impurities is straightforward as,
\begin{gather}
 S_3 \leftarrow \frac{1}{8}
  \left( S_3 T + 3 S_2 S_1 + 3 S_1 S_2 + T S_3
  \right), \\
 S_4 \leftarrow \frac{1}{16}
  \left( S_4 T + 4 S_3 S_1 + 6 S_2 S_2 + 4 S_1 S_3 + T S_4
  \right).
\end{gather}
The higher-order moment of the magnetization, $\langle m^k\rangle$ is
calculated by the ratio of $\operatorname{Tr} S_k$ to $\operatorname{Tr}
T$ similarly to Eq.~(\ref{eq:ave_m}).

Note that we use the same isometric tensor $U^{(t)}$ as for the local
tensor regardless of impurity tensors in order to fix the definition of
bond indices at each level of renormalization.  If we use different
isometries for each impurity tensors, they cannot connect properly in
the next renormalization step.

So far, we focus on the spacial average of a physical quantity defined
at each site but our technique can be applied to a quantity defined on a
small cluster, e.g., the interaction energy defined on each pair of
spins,
\begin{equation}
 e \equiv \frac{1}{N}\sum_{\langle ij\rangle}e_{\sigma_i \sigma_j}.
\end{equation}
To define an impurity tensor for such a impurity located on a bond, it
is useful that a change by impurities is absorbed into the local weight
$W_{k,\sigma\sigma'}\equiv W_{\sigma\sigma'} \left(e_{\sigma\sigma'}\right)^k$.
We define $Y_k\equiv W_k \tilde{U}\Lambda^{-1/2}$ which satisfies $Y_0=X$ and
$Y_k X^\dagger = W_k$.  On these settings, the single and double impurity
tensors with impurities on bonds $x$ or $y$ are written as
\begin{equation}
 E_{1,xyx'y'}^{(0)} =\sum_{\sigma}
  (Y_{1,\sigma x}X_{\sigma y}+ X_{\sigma x}Y_{1,\sigma y})
  X_{\sigma x'}^*X_{\sigma y'}^* V_{\sigma}.
\end{equation}
\begin{multline}
 E_{2,xyx'y'}^{(0)} =\sum_{\sigma}
 (Y_{2,\sigma x}X_{\sigma y} + 2 Y_{1,\sigma x}Y_{1,\sigma y}\\
 + X_{\sigma x}Y_{2,\sigma y})
 X_{\sigma x'}^*X_{\sigma y'}^* V_{\sigma}.
\end{multline}
Its update rule is the same as the magnetization, $S_k$.  Although here
we put no impurity on bonds $x'$ and $y'$, more symmetric definition is
possible.  However, additional treatments are necessary for higher-order
moments if we put impurities on bonds $x'$ and $y'$, since
$Y_k Y_{k'}^\dagger \neq W_{k+k'}$ in general.

The update rule of a tensor with multiple kinds of impurities is
straightforward.  For example, if $S_{k,k'}$ represents a coarse-grained
tensor with $k$ impurities and $k'$ other impurities, we have
\begin{equation}
 S_{1,1} \leftarrow \frac{1}{4}
  (S_{1,1}T + S_{1,0}S_{0,1} + S_{0,1}S_{1,0} + TS_{1,1}).
\end{equation}
To obtain $S_{2,2}$, we need to calculate nine kinds of tensors,
$\{T,S_{1,0}, S_{0,1}, S_{2,0}, S_{1,1}, S_{0,2}, S_{2,1}, S_{1,2},
S_{2,2}\}$,
whose computational cost is 36 times larger than the cost only for $T$.

\section{Numerical results}

The $q$-state Potts model~\cite{Potts,Wu1982} on the square lattice is
defined by the local Boltzmann factors
\begin{equation}
 W_{\sigma,\sigma'} = e^{K \delta_{\sigma,\sigma'}}, \quad
  V_{\sigma} = e^{K' \delta_{\sigma,0}}.
\end{equation}
The spin variable $\sigma$ takes an integral value from $0$ to $q-1$ and
the Kronecker's delta $\delta_{\sigma,\sigma'}$ takes unity if
$\sigma=\sigma'$ and zero otherwise.  The parameters $K$ and $K'$
include the inverse of temperature.  The $q$-state Potts model without
the magnetic field ($K'=0$) exhibits a phase transition at $K_c =
\log(1+\sqrt{q})$.  The phase transition is continuous for $q\leq 4$ and
discontinuous otherwise.  The eigenvalue decomposition of $W$ yields
\begin{equation}
 X_{\sigma x}=\frac{e^{i2\pi \sigma x/q}}{\sqrt{q}} \sqrt{\lambda_x},\label{eq:mat_X}
\end{equation}
where $\lambda_x= e^{K}-1+q\delta_{x,0}$ is the eigenvalue of the
local Boltzmann weight and a bond index $x$ also takes an integer value from
$0$ to $q-1$.

The Potts model without the magnetic field has the global $Z_q$
symmetry, a subgroup of the symmetric group $S_q$.  Therefore the local
tensor $T$ is represented as a $Z_q$-symmetric tensor
\cite{Singh2010,Singh2011}.  After summation over spin degree of
freedom, elements of the local tensor are written as
\begin{equation}
 T_{xyx'y'} = \frac{\sqrt{\lambda_x\lambda_y\lambda_{x'}\lambda_{y'}}}{q}
  \Delta_q(x+y-x'-y'),
\end{equation}
where $\Delta_q(x)$ takes one if $x\equiv 0$ modulo $q$ and zero
otherwise.  A shift of all spins $\sigma \rightarrow \sigma+\tau$ causes
phase factors $e^{i2\pi\tau x/q}$ in Eq.(\ref{eq:mat_X}).  The local tensor
$T$ is clearly invariant under such a phase shift owing to
cancellation by $\Delta_q(x)$.

We consider a complex magnetization as a order parameter defined by
\begin{equation}
 m_\sigma = e^{i2\pi \sigma/q},
\end{equation}
which is suitable for the global $Z_q$ symmetry.
Although the $S_q$ symmetry is spontaneously broken in the
ferromagnetic phase of the Potts model, the above definition is
sufficient to detect the phase transition.  As we will show later, the
corresponding impurity tensors are covariant or invariant under the
$Z_q$ transformation.  However, since the normal second momentum
$\langle m^2 \rangle$ vanishes except the Ising case ($q=2$), we
consider the square of the norm of the magnetization, $\langle |m|^2
\rangle=\langle mm^* \rangle$.  Thus, we treat the complex conjugate of
the local magnetization, $m_\sigma^*$, as another kind of impurities and
use the extended systematic summation scheme as presented in the
previous section.

The initial impurity tensor with $k$ impurities and $k'$
complex-conjugate impurities is written as
\begin{equation}
 S_{k,k';xyx'y'}^{(0)} =
  \frac{\sqrt{\lambda_x\lambda_y\lambda_{x'}\lambda_{y'}}}{q}
  \Delta_q(x+y-x'-y'+k-k').
\end{equation}
This real-valued tensor is invariant under the $Z_q$ transformation if
$k\equiv k'$ (mod $q$) and covariant otherwise, that is, the spin shift
causes only a phase factor,
\begin{equation}
 S_{k,k'}^{(0)} \rightarrow e^{i2\pi\tau (k-k')/q} S_{k,k'}^{(0)}.
\end{equation}
This property holds at any renormalization step if the isometric tensor
$U^{(t)}$ is invariant or covariant under the $Z_q$ transformation.  In
this paper, we have imposed invariance on $U^{(t)}$ without loss of
generality.

The Binder ratio for the complex magnetization is now defined as
\begin{equation}
 U_4 \equiv \frac{\langle |m|^4 \rangle}{ \langle |m|^2 \rangle^2}
  = \frac{\operatorname{Tr} S_{2,2} \operatorname{Tr} T}
  {\left(\operatorname{Tr} S_{1,1}\right)^2}.
\end{equation}
We have, in the high temperature limit ($K\rightarrow 0$),
$U_4\rightarrow 3$ in the Ising model and $U_4\rightarrow 2$ otherwise,
while its low temperature limit ($K\rightarrow\infty$) is
$U_4\rightarrow 1$ regardless of the number of states.  In the
thermodynamic limit, the Binder ratio becomes the step function whose
jump occurs at the transition temperature.

\begin{figure}
 \centering
 \includegraphics[scale=0.38]{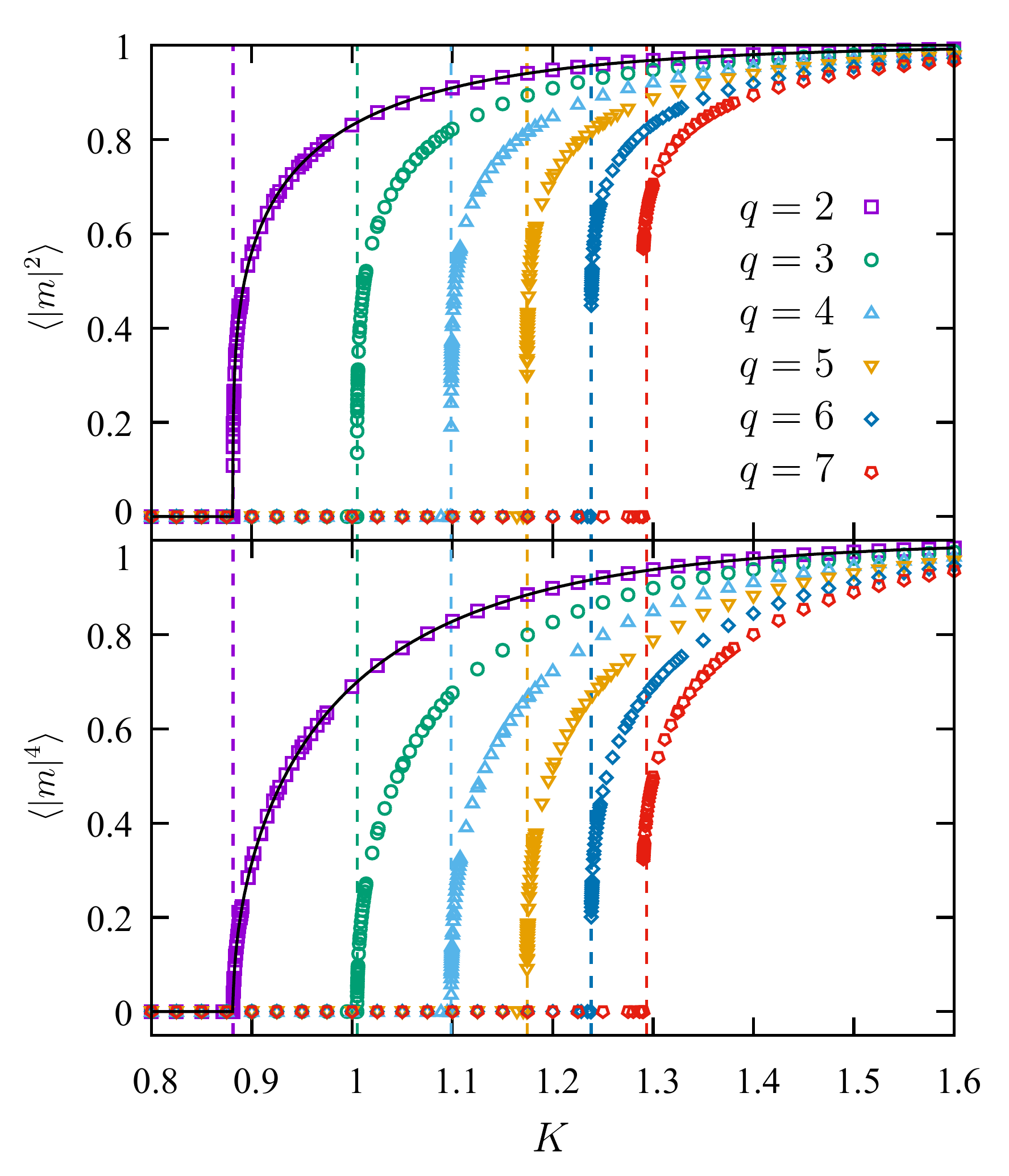}
 \caption{\label{fig:mag_potts} Temperature dependence of the second-
 and fourth-order moments of the magnetization of the $q$-state Potts
 model on a $2^{20}\times 2^{20}$ square lattice obtained by HOTRG with
 $\chi=48$.  The vertical dashed lines indicate the exact transition point
 $K_c$. The solid line shows the exact results of the Ising
 model ($q=2$) in the thermodynamic limit \cite{Yang}.}
\end{figure}

\begin{figure}
 \centering
 \includegraphics[scale=0.38]{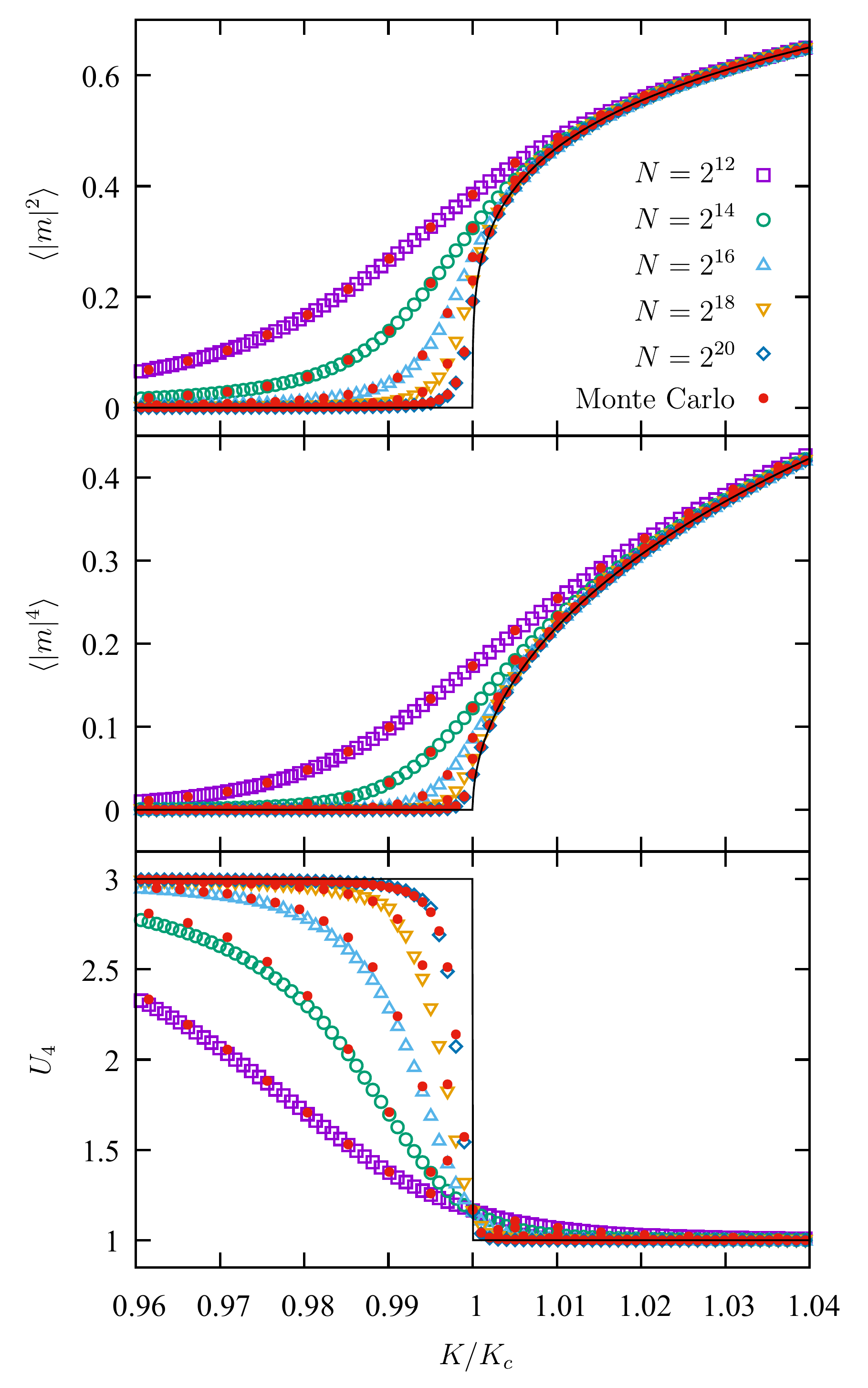}
 \caption{\label{fig:mag_ising} Comparison between HOTRG with $\chi=48$
 and the cluster Monte Carlo (MC) method on the Ising model.  The filled
 circles indicate MC results, whose error bar is smaller than the size
 of a point. The solid line shows the exact results in the thermodynamic
 limit \cite{Yang}.}
\end{figure}

\begin{figure}
 \centering
 \includegraphics[scale=0.4]{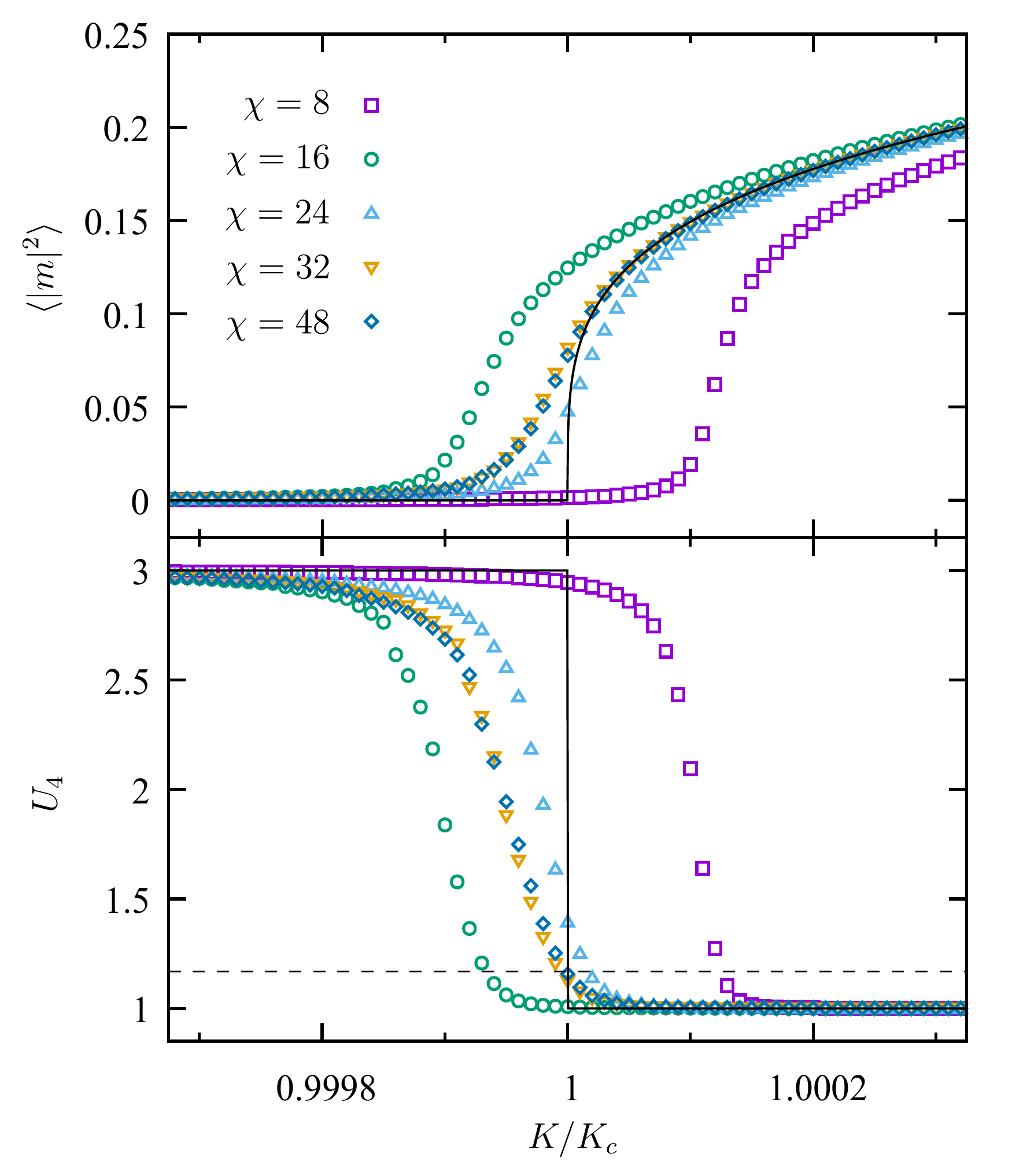}
 \caption{\label{fig:binder_vs_chi} The $\chi$-dependence of the
 magnetization $\langle |m|^2\rangle$ and the Binder ratio $U_4$ after
 the $30$-th HOTRG step ($N=2^{30}$) on the Ising model. The solid line
 shows the exact behavior in the thermodynamic limit \cite{Yang} and the
 horizontal dashed line indicates the critical Binder ratio, taken from
 Ref.~\cite{universal_Binder}. }
\end{figure}

Temperature dependence of the magnetization on the $q$-state Potts model
($q=2,\dots,7$) without the external magnetic field is shown in
Fig.~\ref{fig:mag_potts}.  Here we perform $40$ steps of HOTRG
calculation ($N=2^{40}$) with the bond dimension $\chi=48$.  The HOTRG
results in the Ising model ($q=2$) are on a curve of the exact result in
the thermodynamic limit \cite{Yang}.  Properties of phase transitions will
be discussed in the next section.

Next, we compare the proposed method with the Monte Carlo (MC)
simulations in the Ising model ($q=2$) on small systems
(Fig.\ref{fig:mag_ising}).  We use the Swendsen-Wang algorithm for the
MC simulations~\cite{SwendsenWang}.  The magnetization, $\langle
|m|^2\rangle$ and $\langle |m|^4\rangle$, agree well with the MC
results.  However, the Binder ratio shows slight deviation from the MC
results because of truncation errors in the renormalization.  As shown
in Fig.~\ref{fig:binder_vs_chi}, the Binder ratio with the finite bond
dimension shows a sharp drop whose location fluctuates around the exact
critical temperature and seems to converge to the exact solution in the
limit $\chi\rightarrow\infty$.  At criticality, the Binder ratio takes a
universal value.  For the Ising model, $U_4=1.167929(1)$ is reported in
Ref.~\cite{universal_Binder}, while the HOTRG calculation with $\chi=48$
produces $U_4=1.1574$ after $30$ renormalization steps ($N=2^{30}$).

\begin{figure}
 \centering
 \includegraphics[scale=0.4]{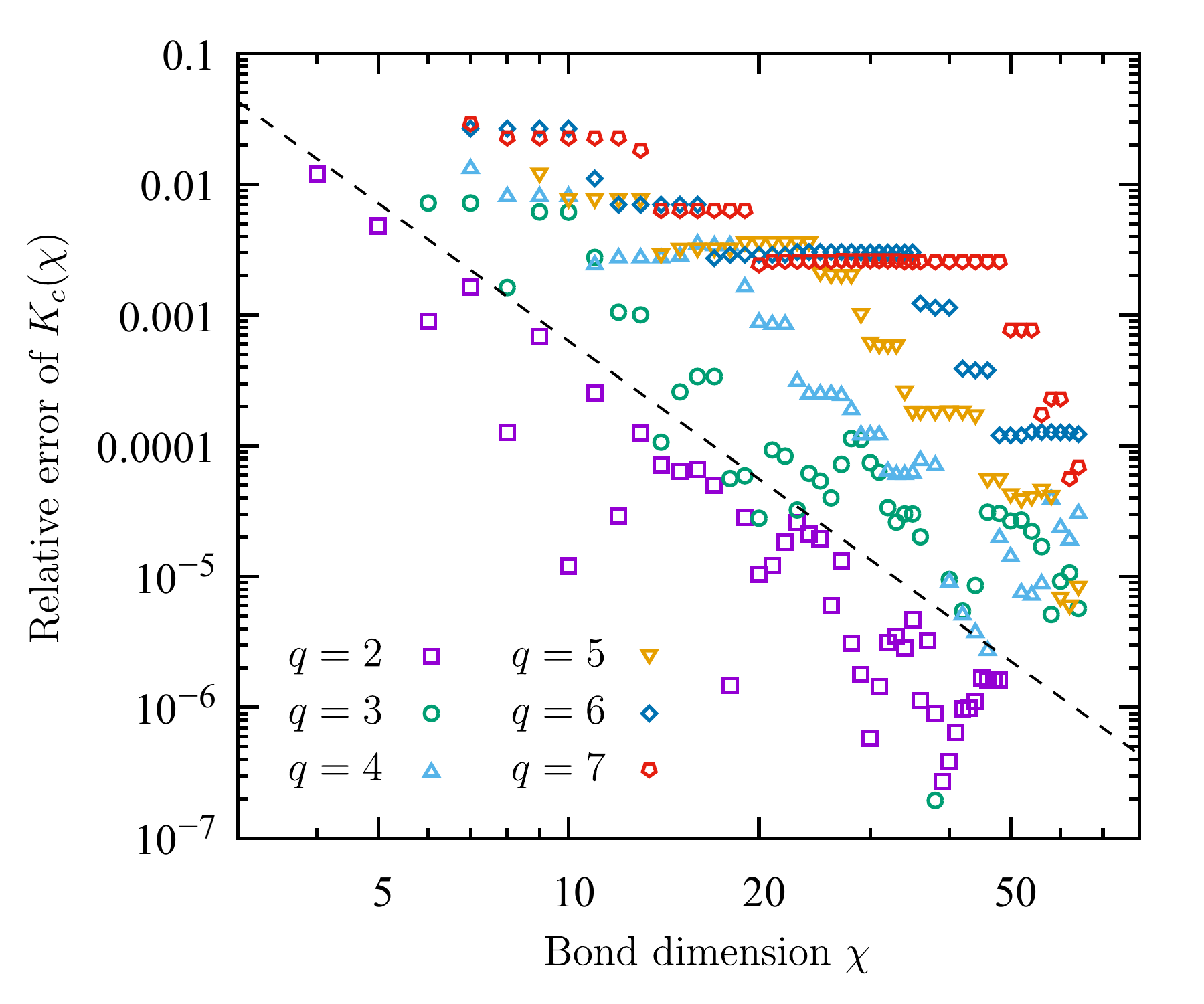}
 \caption{\label{fig:Tc_err} The relative error of the estimated
 transition point $K_c(\chi)$ from the exact solution $K_c$.  The dashed
 line is proportional to $\chi^{-3.5}$.}
\end{figure}

The transition temperature $K_c(\chi)$ estimated under the fixed bond
dimension $\chi$ is determined by a location of a jump of the Binder
ratio.  We search a jump point by using the bisection algorithm.  We
observe that $K_c(\chi)$ oscillates around the exact transition
temperature and converges to $K_c$ in the thermodynamic limit.  Plateaus
of $K_c(\chi)$ for large $q$ are due to degeneracy of the entanglement
spectrum.  The relative error $\Delta K_c \equiv (K_c(\chi)-K_c)/K_c$
seems to decrease with $\chi^{-3.5}$ (Fig.~\ref{fig:Tc_err}), which will
be discussed later.

\section{Finite-size scaling analysis}

\begin{figure}[h]
 \centering
 \includegraphics[scale=0.4]{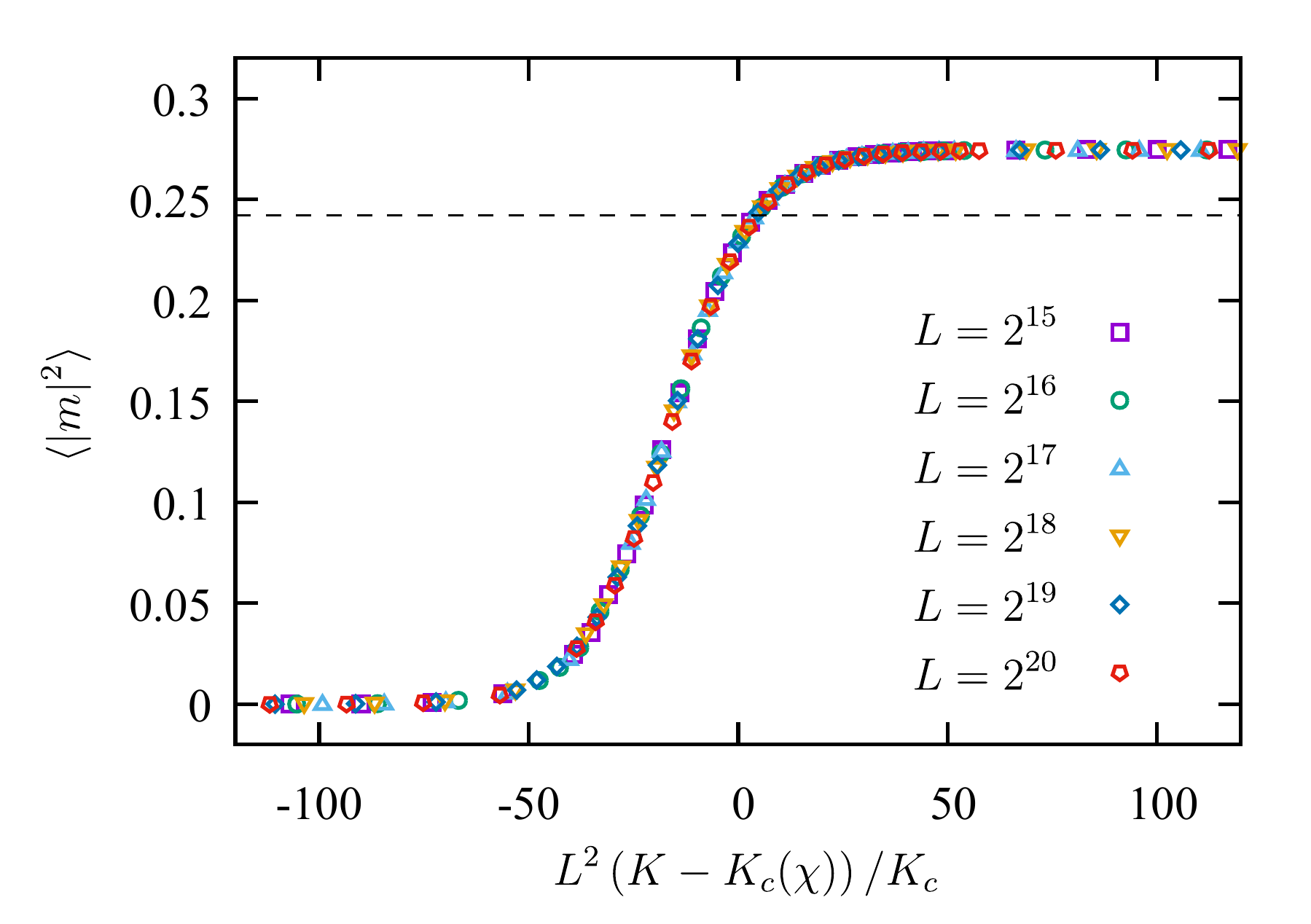}
 \caption{\label{fig:mag2_FSS_Q05} Finite-size scaling plots of the
 magnetization $\langle |m|^2\rangle$ on the 5-state Potts models by
 HOTRG with $\chi=48$. The critical exponents are fixed to the known
 values for the first-order transition, that is, $1/\nu=2$ and
 $\beta=0$.  The horizontal dashed line indicates the exact value of
 spontaneous magnetization at the transition
 temperature~\cite{potts_exact_mag}.}
\end{figure}

The finite-size scaling (FSS) analysis of the magnetization
$\langle|m|^2\rangle$ makes clear distinction between discontinuous and
continuous phase transitions. In the $5$-state Potts model, the phase
transition is weekly first-order and its critical exponent is
$1/\nu=d=2$ and $\beta=0$~\cite{Nienhuis1975,Fisher1982}. The HOTRG
result falls on a single curve by using the standard scaling form
\begin{equation}
 \langle |m|^2 \rangle \sim L^{-2\beta/\nu} g(L^{1/\nu} \delta).
  \label{eq:FSS_mag2}
\end{equation}
with the exact exponents (Fig.~\ref{fig:mag2_FSS_Q05}) where $L$ denotes
the system length ($N=L^2$).  The reduced temperature is defined as
$\delta\equiv(K-K_c(\chi))/K_c$ and we use the estimated critical
temperature $K_c(\chi)$ already obtained from a jump of the Binder
ratio.  The plateau of the scaling function in low-temperature region
also implies discontinuity of the magnetization at the first-order phase
transition point.
The plateau height overestimates the magnetization discontinuity at the
transition point, whose exact value is known as $\Delta
m^2_\text{exact}=0.24220$ for the $5$-state Potts
model~\cite{potts_exact_mag}.  Similar to the estimated transition
temperature $K_c(\chi)$, we observed that the plateau height depends on the
bond dimension and fluctuates around the exact value.

\begin{figure}[h]
 \centering
 \includegraphics[scale=0.4]{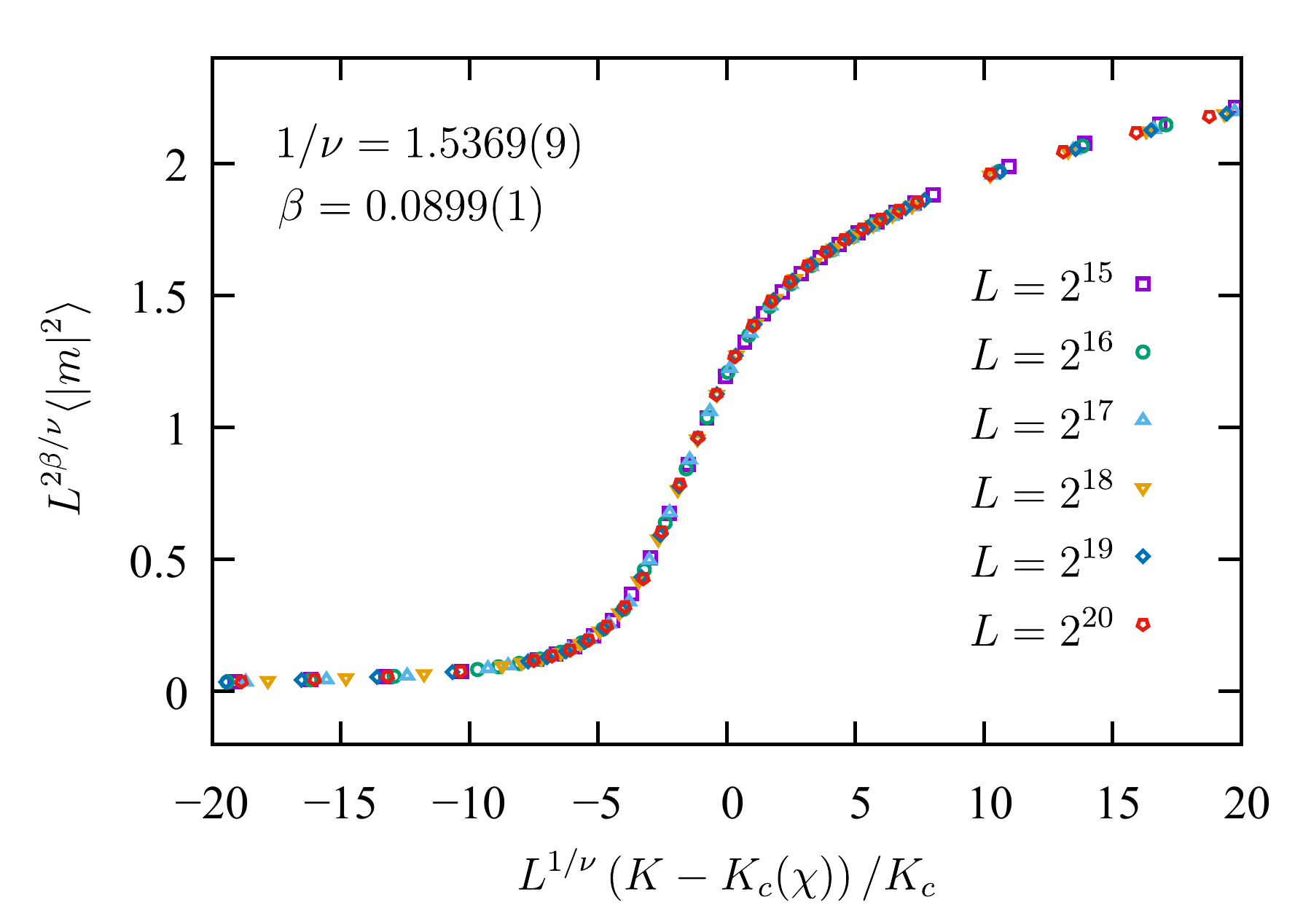}
 \caption{\label{fig:mag2_FSS_Q04} Finite-size scaling plots of the
 magnetization $\langle |m|^2\rangle$ on the 4-state Potts models by
 HOTRG with $\chi=48$.  We assumed the scaling form
 Eq.~(\ref{eq:FSS_mag2}) without the logarithmic multiplicative
 corrections.}
\end{figure}

On the other hand, the phase transition of the $4$-state Potts model is
continuous.  The FSS analysis of the HOTRG result with $\chi=48$ is
shown in Fig.~\ref{fig:mag2_FSS_Q04}.  The critical exponents are
estimated as $1/\nu=1.5369(9)$, $\beta=0.0899(1)$ using the kernel
method proposed in Ref.~\cite{BSA}.  These values are close to the known
exact result $1/\nu=3/2$ and $\beta=1/12=0.0833\cdots$.  During the
FSS procedure, the critical temperature $K_c(\chi)$ is fixed to the
value estimated from a jump of the Binder ratio as mentioned before.  We
need to comment on the fact that the $4$-state Potts model has a
marginal operator which yields logarithmic multiplicative
corrections~\cite{Nauenberg_Scalapino, Cardy_Nauenberg_Scalapino,
Salas_Sokal}.  However, since the system size we used for FSS is
sufficiently large, significant effect for the logarithmic corrections is
not observed.  Assuming the scaling form with the logarithmic corrections
\begin{equation}
 \langle |m|^2 \rangle \sim
  L^{-2\beta/\nu} (\log L)^{1/8} g'(L^{1/\nu} (\log L)^{-3/4} t),
\end{equation}
we estimate critical exponents as $1/\nu=1.5992(8)$ and
$\beta=0.0832(1)$.

\begin{figure}
 \centering
 \includegraphics[scale=0.4]{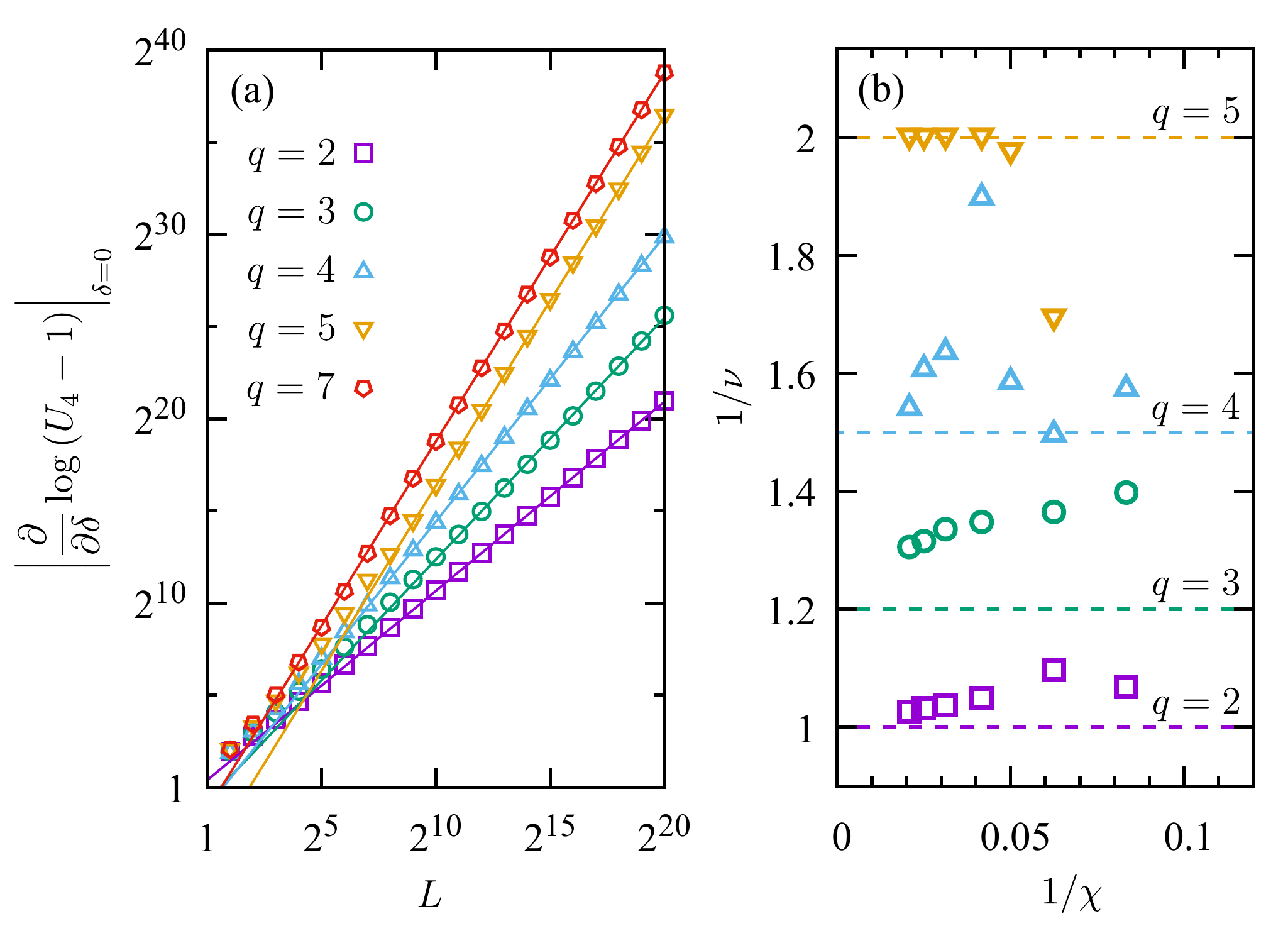}
 \caption{\label{fig:Binder_log_deriv} (a) Derivative of $\log(U_4-1)$
 at the estimated transition point with $\chi=48$.  (b) The
 $\chi$-dependence of the estimated critical exponents $1/\nu$.  The
 dashed horizontal lines indicate the exact values.}
\end{figure}

A slope of physical quantities at the transition temperature also
provides information about critical exponents. Especially the Binder
ratio, a dimensionless quantity, satisfies
\begin{equation}
 \log \left| \frac{\partial U_4}{\partial \delta} \right|_{\delta=0}
  = c + \frac{1}{\nu} \log L + O(L^{-\omega}),\label{eq:log_deriv_U4}
\end{equation}
where $\omega$ is an exponent for correction to scaling.  Since we
already precisely estimated the transition point with fixed bond
dimension from a jump of the Binder ratio, we use a definition
$\delta\equiv(K-K_c(\chi))/K_c$.  It is different from the crossing point
analysis for the Monte Carlo simulation~\cite{Shao_Guo_Sandvik}.  The
left-hand side of Eq.~(\ref{eq:log_deriv_U4}) fits well to a linear line
for large system size (Fig.~\ref{fig:Binder_log_deriv}(a)).
Here, we consider $\log(U_4-1)$ instead of $U_4$ since it is linear
with respect to the inverse temperature around the first-order
transition point. (Its phenomenological argument was shown in
Ref.~\cite{iino2018}.)  The system size where the data are on a line is
smaller than the known correlation length.  The correlation length at
the transition point approaching from the disordered phase is 2512.2 and
48.1 for $q=5$ and $7$, respectively \cite{potts_cor_length}.  In the
$5$-state Potts model, a point at $L=512$ is already on the fitting
line. This behavior is consistent with the finite-size scaling analysis
reported in Ref.~\cite{iino2018}.

\begin{table*}
 \centering
 \caption{\label{table:nu} The critical exponents $1/\nu$ estimated from
 a slope of the Binder ratio with $\chi=48$.  An error indicates the
 standard error of the fit.}
 \begin{tabular}{c|ccccc} \hline
   $q$ & 2 & 3 & 4 & 5 & 7\\ \hline
   Exact & 1 & $6/5$ & $3/2$ & 2 & 2\\
   Slope & 1.026(2) & 1.305(10) & 1.544(1) & 2.006(2) & 2.0001(5) \\ \hline
 \end{tabular}
\end{table*}

The estimated critical exponent $1/\nu$ seems to converge to the exact
value in the limit $\chi\rightarrow\infty$
(Fig.~\ref{fig:Binder_log_deriv}(b)).  Large correction to scaling may
cause instability in $q=4$ in contrast to FSS of the squared
magnetization.  In Table~\ref{table:nu}, we list the critical exponent
$1/\nu$ estimated by HOTRG calculations with $\chi=48$.  An effect of
the finite bond dimension is larger than the standard error of the fit.
Our method provides less accurate critical exponent than the CTMRG
method~\cite{CTMRG1997}. However we would like to emphasize that our
analysis does not use any assumptions about the exact transition
temperature and the exact energy at criticality, while the CTMRG result
used them.  Though using the exact values usually increases the accuracy
of the estimates drastically, they are not available in general
circumstances.  The scaling dimensions can also be determined from the
eigenvalue of the transfer matrix~\cite{TEFR2009}.  However, they are
not stable in the HOTRG procedure because of residual short-range
correlation as well as TRG~\cite{loopTNR2017}.

\section{Conclusions and Discussions}

In this paper, we have proposed an algorithm for the higher-order
moments of physical quantities based on the HOTRG method.  The
systematic summation technique provides the recursion formula for
coarse-graining impurity tensors. We have shown that this method
calculates the order parameter and the Binder ratio in the Potts model
on the square lattice.  The transition temperature is precisely determined
from a jump of the Binder ratio.  Our results are more accurate than the
Monte Carlo results. For example, we have achieved $\Delta K_c =
8.3\times 10^{-6}$ in the 5-state Potts model, while the nonequilibrium
relaxation analysis did $2.5\times 10^{-5}$~\cite{Ozeki2003}.  The
critical exponents are reproduced via FSS analysis.  The weakly
first-order transition in the $5$-state Potts model is clearly distinct
from the continuous transition.  This judgment has been made possible by
the method proposed in this article. There are two essential points: (i)
the advantage in the system size compared to the MC approaches, and (ii)
the advantage of symmetry-respecting method over the numerical
differentiation approaches.

Our method easily approaches a huge system size which the MC method
cannot treat.  In the case of no magnetic field, our scheme does not break
the spin-rotational symmetry, in other words, each tensor is invariant
or covariant under the spin rotation.  This fact stabilizes calculations
and is an advantage over the numerical differentiation approach because
the latter needs to introduce a tiny magnetic field which breaks
symmetry.  Although we considered the two-dimensional square lattice in
this paper, generalization to higher-dimensional systems is
straightforward as well as the HOTRG method.

The accuracy of the HOTRG method depends on the bond dimension $\chi$.
We observed that the shift of effective critical point roughly scaled as
$\Delta K_c \sim \chi^{-3.5}$ (Fig.~\ref{fig:Tc_err}).  This behavior
would relate to the finite-$\chi$ scaling.  The effective correlation
length satisfies $\xi_\text{eff}\sim\chi^\kappa$ and the argument of the
entanglement entropy based on the conformal invariance provides
\begin{equation}
 \kappa = \frac{6}{c\left(\sqrt{12/c}+1\right)},
\end{equation}
where $c$ is the central charge.  For the Ising model ($c=1/2$),
three-state ($c=4/5$) and four-state ($c=1$) Potts models, we have
$\kappa=2.034$, $1.539$, and $1.334$, respectively.  On the other hand,
the scaling of the effective correlation length implies $\Delta K_c \sim
\chi^{-\kappa/\nu}$.  However, our observation is inconsistent with this
behavior. It indicates that the naive scaling relation for $\Delta K_c$
needs to be correct, which remains as a future issue.

Lastly, we would like to mention possibilities of further improvements
in the present method.  One is an introduction of the environment
tensor. The higher-order second renormalization group method successes to
significantly improve the accuracy by globally optimizing the truncation
scheme~\cite{HOTRG2012}.
The other is entanglement filtering~\cite{TEFR2009}.  It is known that
HOTRG does not remove short-range entanglement completely and eventually
gets stuck into a fictitious fixed point~\cite{Ueda2014}.  Entanglement
filtering techniques, such as a loop-TNR approach~\cite{loopTNR2017}, a
graph independent local truncation's (GILT)~\cite{GILT}, and a full
environment truncation (FET)~\cite{FET}, can filter out internal
entanglement.  Recently entanglement branching was proposed as another
approach to manage flow of entanglement~\cite{Branching}.  Thus a
combination of HOTRG with these techniques will catch the true critical
phenomena, which will be considered in future works.

\section*{Acknowledgments}
The authors would like to thank K.~Harada, T.~Okubo, H.~Ueda,
Y.~Motoyama, S.~Iino and H.~Watanabe for valuable discussions.  The
computation in the present work is partially executed on computers at
the Supercomputer Center, ISSP, University of Tokyo.  This research was
supported by MEXT as "Exploratory Challenge on Post-K computer"
(Frontiers of Basic Science: Challenging the Limits), and by ImPACT
Program of Council for Science, Technology and Innovation (Cabinet
Office, Government of Japan).

\section*{References}
\bibliography{main}

\end{document}